\pdfoutput=1  % for PDFLaTeX compilation by arXiv
\documentclass{JFMtemplate/jfm}
\usepackage{graphicx}
\usepackage{amsmath}
\usepackage{gensymb}  % for \degree

\renewcommand{\vec}{\boldsymbol}  % bold vectors
\newcommand{\Ray}{\mbox{\textit{Ra}}}  % Rayleigh number
\newcommand{\Nuss}{\mbox{\textit{Nu}}}  % Nusselt number

\title{Combined measurement of velocity and temperature in liquid metal convection}
\shorttitle{Combined measurement of velocity and temperature in liquid metal convection}

\author{
  Till~Z\"urner\aff{1}
  \corresp{\email{till.zuerner@tu-ilmenau.de}},
  Felix~Schindler\aff{2},
  Tobias~Vogt\aff{2},
  Sven~Eckert\aff{2}
  \and
  J\"org~Schumacher\aff{1}}
\shortauthor{
  T. Z\"urner, F. Schindler, T. Vogt, S. Eckert and J. Schumacher}

\affiliation{
\aff{1}Institute of Thermodynamics and Fluid Mechanics, Technische Universit\"at Ilmenau, Postfach~100565, D-98684 Ilmenau, Germany
\aff{2}Department of Magnetohydrodynamics, Institute of Fluid Dynamics, Helmholtz-Zentrum Dresden -- Rossendorf, Bautzner Landstra\ss e 400, D-01328 Dresden, Germany}

\begin{document}

\maketitle

\begin{abstract}
Combined measurements of velocity components and temperature in a turbulent Rayleigh-B\'enard convection flow at a low Prandtl number of $\Pran = 0.029$ and Rayleigh numbers between $10^6\le \Ray\le 6\times 10^7$ are conducted in a series of experiments with durations of more than a thousand free-fall time units.
Multiple crossing ultrasound beam lines and an array of thermocouples at mid-height allow for a detailed analysis and characterization of the complex three-dimensional dynamics of the single large-scale circulation (LSC) roll in the cylindrical convection cell of unit aspect ratio which is filled with the liquid metal alloy GaInSn.
We measure the internal temporal correlations of the complex large-scale flow and distinguish between short-term oscillations associated with a sloshing motion in the mid-plane as well as varying orientation angles of the velocity close to the top/bottom plates and the slow azimuthal drift of the mean orientation of the roll as a whole that proceeds on an up to a hundred times slower time scale. 
The coherent LSC drives a vigorous turbulence in the whole cell that is quantified by direct Reynolds number measurements at different locations in the cell.
The velocity increment statistics in the bulk of the cell displays characteristic properties of intermittent small-scale fluid turbulence. 
We also show that the impact of the symmetry-breaking large-scale flow persists to small-scale velocity fluctuations thus preventing the establishment of fully isotropic turbulence in the cell centre. 
Reynolds number amplitudes depend sensitively on beam line position in the cell such that different definitions have to be compared. 
The global momentum and heat transfer scalings with Rayleigh number are found to agree with those of direct numerical simulations and other laboratory experiments.
\end{abstract}

\section{Introduction}
\label{sec:intro}

The understanding of transport processes in several turbulent convection flows in nature and technology can be improved by means of Rayleigh-B\'enard convection (RBC) studies at very low Prandtl numbers of $\Pran \ll 10^{-1}$.
Prominent examples are stellar and solar convection \citep{Spiegel1962}, the geodynamo in the core of the Earth \citep{Christensen2006}, the blanket design in nuclear fusion reactors \citep{Salavy2007} or liquid metal batteries for renewable energy storage \citep{Kelley2018}.
Laboratory experiments in turbulent RBC at low Prandtl numbers are, however, notoriously challenging since they have to rely on liquid metals as working fluid to obtain a sufficiently high thermal diffusivity in comparison to the kinematic viscosity.
Liquid metals are opaque and thus exclude optical imaging by means of particle image velocimetry \citep{Adrian2011} or Lagrangian particle tracking \citep{Hoyer2005, Toschi2009}.
The analysis relies instead on ultrasound Doppler velocimetry \citep[UDV,][]{Takeda1987} in combination with local temperature measurements.
We mention here pioneering experiments by \citet{Takeshita1996, Cioni1997, Mashiko2004, Tsuji2005} and more recently by \cite{Khalilov2018}, or \citet{Vogt2018a}, who found a jump-rope-type large-scale flow.

In closed convection cells, a large-scale circulation (LSC) builds up that affects the way and amount of heat and momentum carried across the turbulent fluid \citep{Ahlers2009,Chilla2012}.
Its complex three-dimensional shape and dynamics have been studied intensively in the past decade for RBC flows with $\Pran > 0.1$, for example in theoretical oscillator models \citep{Brown2009}, experiments \citep{Funfschilling2004, Sun2005, Xi2009, Zhou2009}, and direct numerical simulations \citep{Stevens2011,Shi2012}.
The low-Prandtl-number regime has been largely unexplored with respect to the large-scale flow dynamics.
Only recently, the interest in this research topic has increased with new experiments by \citet{Khalilov2018} and \citet{Vogt2018a}.
The typical cylindrical cell shape leads to a statistical symmetry of convective turbulence with respect to the azimuthal direction and opens the possibility to complex LSC dynamics.
These consist of shorter-term oscillations of the mean flow orientation close to the plates which point into different directions at top and bottom.
The oscillations can be superimposed by a slow azimuthal drift of the mean flow orientation of the LSC roll as a whole.
The UDV technique has been proved to detect complex flow structures in liquid metal thermal convection \citep{Mashiko2004, Tsuji2005, Vogt2018a, Vogt2018}.
This method has been extended recently to linear transducer arrays that allow to reconstruct 2D flow patterns at high spatial and temporal resolution \citep{Franke2013}.

In this work, we report multi-technique long-term measurements of a fully turbulent convection flow in the liquid metal alloy gallium-indium-tin (GaInSn, $\Pran=0.029$) in a closed cylindrical cell of aspect ratio~1.
We combine 10~UDV beam lines and 11~thermocouples for an in-depth analysis of the LSC at Rayleigh numbers $\Ray\le 6\times 10^7$.
Multiple crossing UDV beam lines close to the bottom/top walls and at the mid-plane in combination with an array of thermocouple probes arranged in a semicircle of high angular resolution at half height enable the detailed experimental reconstruction of a short-term oscillatory torsional motion of the LSC at top and bottom, the sloshing motion at half height and the superposition of this short-term dynamics with a slow azimuthal drift.
The LSC flow is found to be more coherent as in comparable RBC flows at higher Prandtl numbers, agreeing also with recent direct numerical simulations~(DNS) by \citet{Scheel2016,Scheel2017} in the same parameter range.
Our analysis reveals a LSC roll with large inertia, able to drive a vigorous fluid turbulence in the bulk.
This is motivated by a recent DNS study where the higher inertia of fluid turbulence in low-Prandtl-number fluids are found to be largely caused by the injection of turbulent kinetic energy at a larger scale due to the coarser thermal plumes comprising the LSC~\citep{Schumacher2015}.
A reverse influence of the small-scale turbulence on the large-scale flow is, however, also possible as extreme dissipation events may trigger LSC re-orientations~\citep{Schumacher2016a}.
We investigate the turbulent character of the flow from direct determination of the Reynolds number dependence $\Rey(\Ray)$ in the cell centre using UDV.
On the basis of this measurement method, we will also analyse the statistics of velocity increments and the isotropy of small-scale turbulence in the liquid metal flow.
Our experiment yields time series of velocity components and temperature of almost two thousand free-fall times units.
Although three-dimensional high-resolution DNS of such flows provide the full information of the turbulent fields, they cannot be run for extended time intervals of a few hundred free-fall time units or even more \citep{vanderPoel2013,Scheel2016,Scheel2017}.
Laboratory experiments are currently the only way to conduct a long-term global analysis of three-dimensional LSC flow that has to be considered as a superposition of different modes with different typical time scales.

The outline of the paper is as follows. 
In section~\ref{sec:setup} we present details on the experiment. 
Section~\ref{sec:LSC} is dedicated to the large-scale flow in the cell. 
We discuss the oscillation of the azimuthal orientation, the torsion as well as the sloshing modes by means of three representative runs at $\Ray = 10^6$, $10^7$ and $6\times 10^7$. 
For the latter, we study a cessation event in detail. 
Furthermore, in this section we take a closer look to the internal temporal correlations of the large-scale flow. 
Section~\ref{sec:transport} reviews the global transport of heat and momentum. 
Section~\ref{sec:result_turbulence} provides our findings for the statistics of the velocity increments. 
We close the work with a final summary and give a brief outlook into future work.
The values of selected quantities of the presented experiments are published in a supplementary information to this article.

\section{Experimental set-up}
\label{sec:setup}

\begin{figure}
\centerline{
  \includegraphics{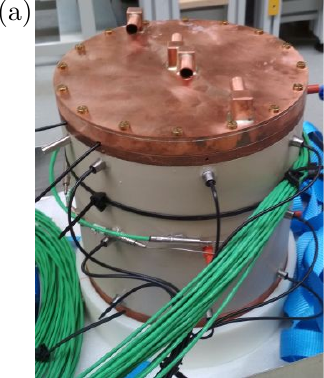} \hfill
  \includegraphics{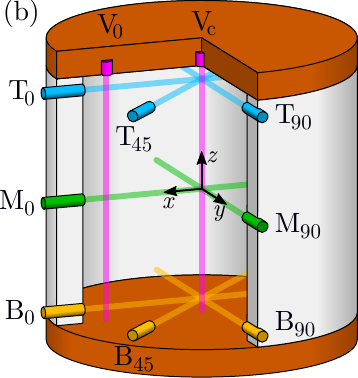} \hfill
  \includegraphics{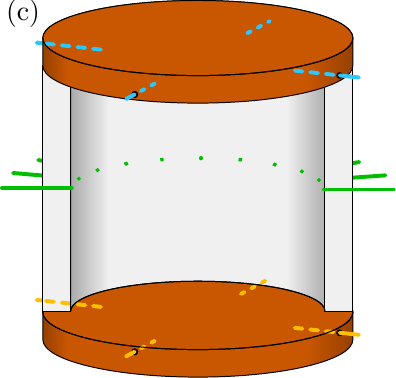}}
\caption{%
  Convection cell design and arrangement of ultrasound beam lines and thermocouples.
  (a)~Photograph.
  (b)~Positions and labels of the ultrasound Doppler velocimetry sensors.
  The subscripts of the labels refer to the azimuthal position in degree.
  (c)~Positions of the temperature sensors.
  The cylindrical convection cell is thermally insulated during the experiment.}
\label{fig:setup_cell}
\end{figure}

The cylindrical convection cell (figure~\ref{fig:setup_cell}(a)) has an inner diameter $D = 2R = 180$\,mm and an inner height $H = 180$\,mm with aspect ratio~$\Gamma = D/H = 1$.
The side walls are made of polyether ether ketone (PEEK), and the top and bottom plates consist of copper with a thickness of 25\,mm.
The top plate is cooled with water supplied by a thermostat.
A ceramic heating plate with a diameter of 190\,mm is mounted below the bottom copper plate, supplying a maximum heating power of 2\,kW for a DC voltage of 230\,V.
The cell is filled with the eutectic alloy GaInSn (melting point $10.5$\,\degree C).
At a mean temperature~$\bar T$ of 35\,\degree C, the melt has a Prandtl number of $\Pran = 0.029$.
The properties of GaInSn are then: mass density $\rho=6.3\times 10^3$\,kg/m$^3$, kinematic viscosity $\nu=3.2\times 10^{-7}$\,m$^2$/s, thermal diffusivity $\kappa=1.1\times 10^{-5}$\,m$^2$/s, thermal conductivity $\lambda=24.3$\,W/(K\,m), and volumetric expansion coefficient $\alpha =1.2\times 10^{-4}$\,K$^{-1}$ \citep{Muller2001, Plevachuk2014}.
The Rayleigh number varies between $10^6\le \Ray\le 6\times 10^7$ thus covering almost two orders of magnitude.
The mean fluid temperature is held at about 35\,\degree C for all experiments, except for the highest temperature differences at $\Ray\sim6\times10^7$.
There, due to limited cooling power, $\bar T$ rises to 40\,\degree C resulting in $\Pran = 0.028$.

The velocity measurements rely on the pulsed UDV technique applying a specific configuration (figure~\ref{fig:setup_cell}(b)), where ten transducers emit ultrasonic pulses with a frequency of 8\,MHz along a straight beam line and record the echoes that are reflected by small particles in the fluid.
Knowing the speed of sound in GaInSn allows us to determine the spatial particle position along the ultrasound propagation from the detected time delay between the burst emission and the echo reception.
The movement of the scattering particles, which are always present in a GaInSn melt, results in a small time shift of the signal structure between two successive bursts from which the velocity can be calculated.

The plate temperatures are measured as the average of four K-thermocouples distributed in 90\degree{} intervals around the plate circumference (dashed lines in figure~\ref{fig:setup_cell}(c)).
Eleven additional thermocouples are arranged at half height of the convection cell in a semicircle in steps of 18\degree{} (solid lines in figure~\ref{fig:setup_cell}(c)).
They span the azimuthal interval of $157.5\degree < \phi < 337.5\degree$, where $\phi=0\degree$ is the positive $x$-axis (see figure~\ref{fig:setup_cell}(b) and~\ref{fig:UDVangle}).

The global heat flux~$\dot Q$ is determined at the top plate:
The inflowing cooling water of temperature $T_\mathrm{in}$ heats up to a temperature $T_\mathrm{out}$ at the outlet.
In combination with the water volume flux $\dot V$ the extracted heat flux is $\dot Q_\mathrm{cool} = \tilde c_p \tilde \rho \dot V (T_\mathrm{out}-T_\mathrm{in})$ with $\tilde c_p$ and $\tilde\rho$ being the isobaric heat capacity and mass density, respectively, of the cooling water at its mean temperature $(T_\mathrm{in}+T_\mathrm{out})/2$~\citep{Cengel2008}.
Only measurements with $T_\mathrm{out}-T_\mathrm{in} > 0.2$\,K are considered due to the incertitude of the thermocouples.

Heat losses to the environment are minimized by insulating the experiment with Styrofoam and placing it on a polyamide base with a thermal conductivity $\sim 0.23$\,W/(m\,K).
Any remaining heat losses through the side wall are determined by measuring the radial temperature gradient~$\partial_r T$ within the side wall. 
Three pairs of thermocouples are distributed in $120$\degree{} intervals around the circumference at half-height of the cell.
The vertical position is taken as representative of the average heat loss over the cell height and the three azimuthal positions are to minimize anisotropic effects caused by the LSC.
The radial heat flux is given by $\dot Q_\mathrm{loss} = -\lambda_\mathrm{sidewall} \pi DH \partial_r T$ with $\partial_r T$ as the average value of all three measurement positions, $\pi D H$ as the inner side wall surface, and $\lambda_\mathrm{sidewall}=0.25$\,W/(K\,m).
The calculated heat loss~$\dot Q_\mathrm{loss}$ is, ideally, the difference between the heating power~$\dot Q_\mathrm{heat}$ supplied at the bottom and the cooling power extracted at the top of the cell $\dot Q_\mathrm{loss} = \dot Q_\mathrm{heat} - \dot Q_\mathrm{cool}$.
To calculate the average heat flux~$\dot Q = (\dot Q_\mathrm{heat} + \dot Q_\mathrm{cool})/2$ through the cell, the measured cooling power is corrected by half of the radial heat losses: $\dot Q = \dot Q_\mathrm{cool} + \dot Q_\mathrm{loss}/2$.

\section{Large-scale flow dynamics}
\label{sec:LSC}

\subsection{Long-term dynamics of the large-scale flow structure}
\label{sec:LSC_structure}
%---------------------------------------------------------------
\begin{figure}
\centerline{\includegraphics{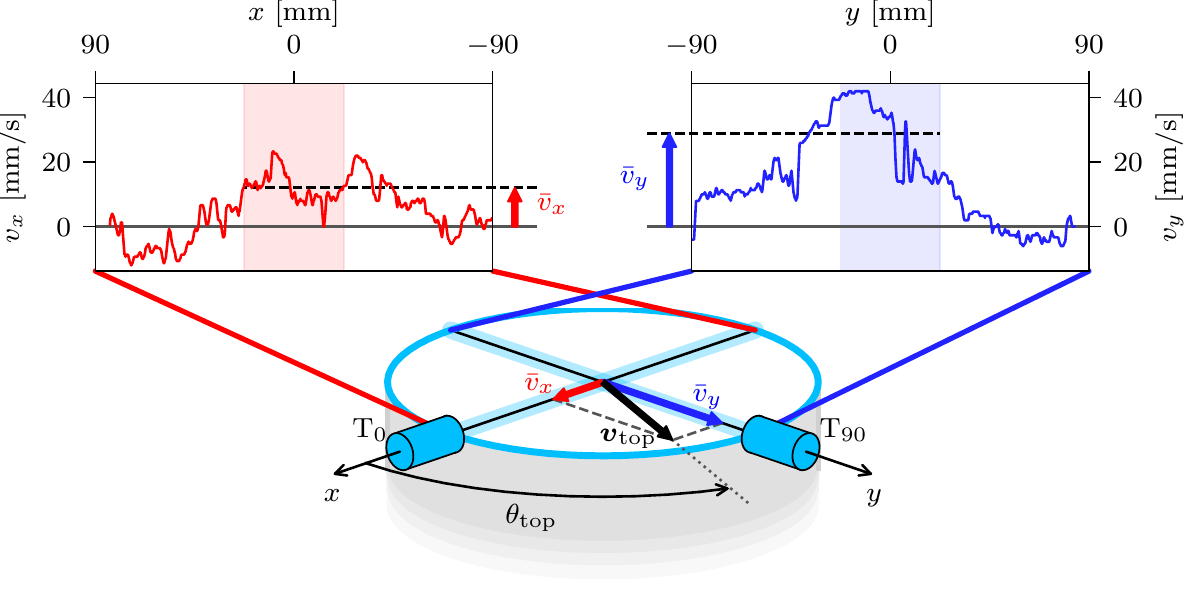}}
\caption{%
  Calculation of the orientation angle $\theta_\mathrm{top}$ of the large-scale flow close to the top plate.
  The UDV sensor T\textsubscript{0} at $\phi=0$\degree{} measures the radial $v_x$ profile (left) and T\textsubscript{90} at $\phi=90$\degree{} measures the radial $v_y$ profile (right).
  The mean velocities $\bar v_x$ and $\bar v_y$ are determined over the central shaded profile intervals.
  These values are used as components of the horizontal velocity vector of the LSC which is denoted as $\vec v_\mathrm{top}$.}
\label{fig:UDVangle}
\end{figure}
%---------------------------------------------------------------

We now discuss the long-time evolution of the LSC for a measurement at our highest Rayleigh number $\Ray=6\times 10^7$.
Times are given in units of the free-fall time $\tau_\mathrm{ff} = \sqrt{H/(g\alpha\Delta T)}$.
Variable $g$ is the acceleration due to gravity.
For the example data set at $\Ray=6\times 10^7$, the free-fall time is $\tau_\mathrm{ff} = 2.3$\,s and the total duration of the time series is $1700 \tau_\mathrm{ff}$.
Figure~\ref{fig:UDVangle} illustrates how the orientation angle of the flow in the centre close to the top plate is calculated.
The velocity profiles $v_x(x,t)|_{y=0}$ and $v_y(y,t)|_{x=0}$ are measured by UDV sensors T\textsubscript{0} and T\textsubscript{90}, respectively, 10\,mm below the top plate.
At the central crossing point of the ultrasonic beams the horizontal velocity vector $\vec v_\mathrm{top} = (\bar v_x, \bar v_y)$ and the resulting orientation angle are given by 
%----------------------------------------------------------------
\begin{equation}
\label{eq:LSCvelocomp}
\bar v_i (t) = \left\langle v_i(i,t) \right\rangle_{-D/8 \le i \le D/8} \,,
\quad\quad\mbox{and thus}\quad\quad
\theta_\mathrm{top} = \arctan\left(\frac{\bar v_y}{\bar v_x}\right) \,,
\end{equation}
%---------------------------------------------------------------
with $i=x,y$.
The orientation angle at the bottom plate $\theta_\mathrm{bot}$ is calculated analogously using UDV sensors B\textsubscript{0} and B\textsubscript{90}.
%----------------------------------------------------------------
\begin{figure}
\centerline{\includegraphics{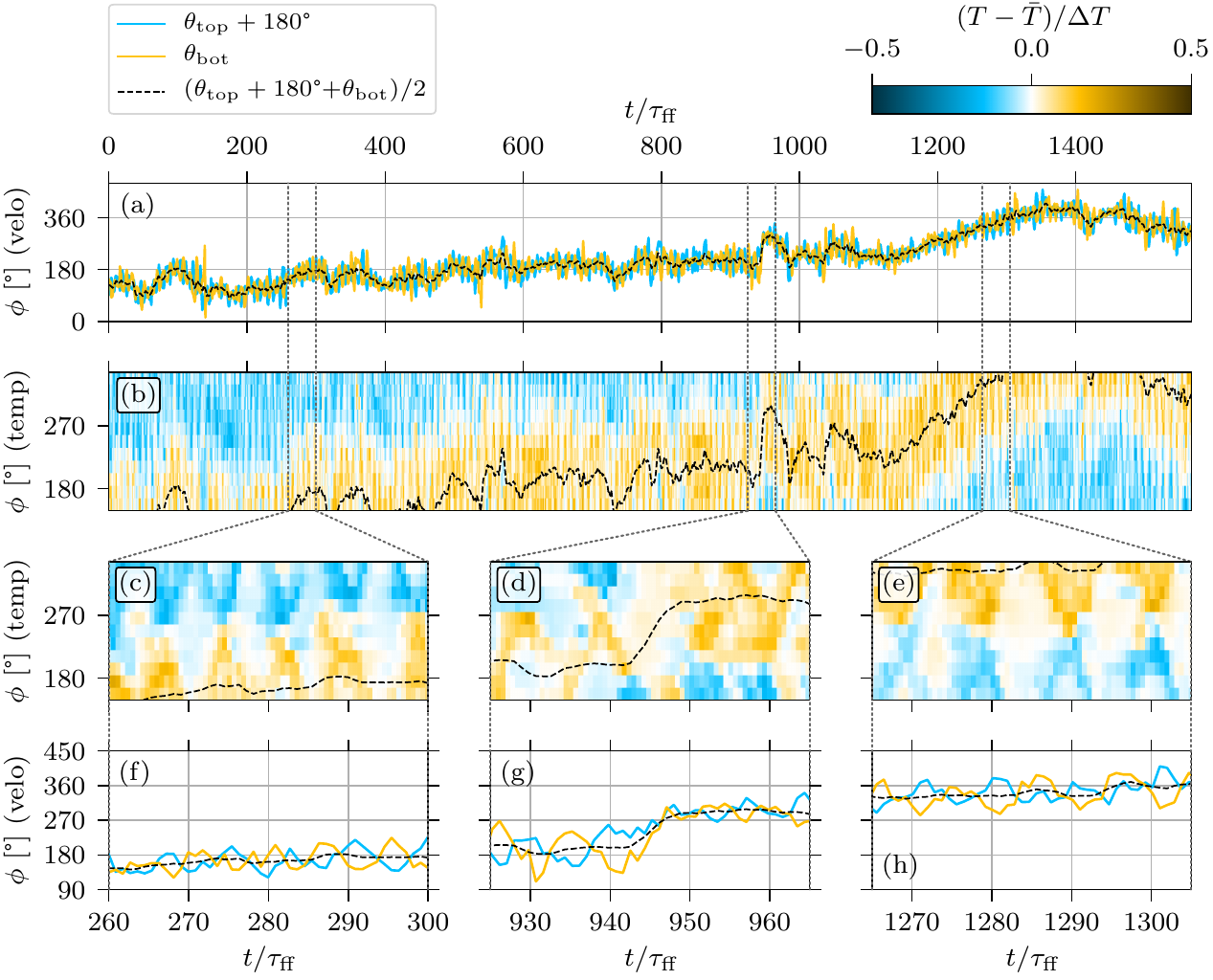}}
\caption{%
  Large-scale flow structure for $\Ray = 6\times10^7$ ($\tau_\mathrm{ff} = 2.3$\,s) measured by UDV sensors near the plates (velo) and thermocouples at mid height (temp).
  (a)~LSC orientation angles at the top ($\theta_\mathrm{top}$) and bottom plate ($\theta_\mathrm{bot}$)  (see figure~\ref{fig:UDVangle}).
  The black dashed line is the smoothed average of the top and bottom angles.
  (b)~Temperatures near the wall at half height of the cell over time and azimuthal position~$\phi$.
  The dashed black line is identical to the one in panel (a).
  (c)--(e):~Detailed views of time series (b) for a time interval of $40\tau_\mathrm{ff}$.
  (f)--(h):~Detailed views of time series (a) for same time interval.
  }
\label{fig:Tmid_angle}
\end{figure}
%----------------------------------------------------------------
Figure~\ref{fig:Tmid_angle}(a) shows a time series of $\theta_\mathrm{top}+180$\degree{} and $\theta_\mathrm{bot}$.
The two angles match very well which implies that in this measurement they always maintain a mean azimuthal offset of 180\degree{}.
This validates the presence of a single coherent LSC roll in the cell.
Figures~\ref{fig:Tmid_angle}(f) and (h) present a detailed view of the data in figure~\ref{fig:Tmid_angle}(a).
It can now be seen that the orientation angles oscillate around a common mean at an oscillation time of $\tau_\mathrm{osc}\sim 10 \tau_\mathrm{ff}$.
Furthermore, the angles oscillate in anti-phase, which is a clear indication of the torsion mode~\citep{Funfschilling2004, Xie2013, Khalilov2018}.
The LSC flow is thus characterized by a torsion in agreement with DNS at similar $\Pran$ by \citet{Scheel2016,Scheel2017} and experiments in liquid sodium~\citep[$\Pran = 0.0094$]{Khalilov2018}.
We come back to this point in section~\ref{sec:LSC_oscfreq}.
Panel~(a) of this figure shows clearly the additional slow drift of the mean orientation angle by more than 180\degree{} over the full measurement time period of $\tau_\mathrm{total}\approx 1700 \tau_\mathrm{ff}$, a motion that is due to the statistical azimuthal symmetry in the cylindrical set-up.
The characteristic time scale~$\tau_\mathrm{drift}$ of this motion can be estimated to be of the order of the thermal diffusion time scale~$\tau_\mathrm{drift} \sim \tau_\kappa = H^2/\kappa$. 
Using $\tau_\kappa = \sqrt{\Ray\Pran}\, \tau_\mathrm{ff}$, this can be expressed in terms of the free-fall time to be $\tau_\mathrm{drift} \sim 170$ to $1300\, \tau_\mathrm{ff}$ for our $\Ray$ range of $10^6 < \Ray < 6 \times 10^7$. 
A quantitative measurement of this time scale would require a Fourier time series analysis of even longer data records than the ones we could obtain.

Most experiments on the large scale flow structure in turbulent RBC use temperature measurements close to the side walls of the cell to infer the azimuthal profile of the vertical flow direction from their temperature imprint:
Up-welling fluid from the hot bottom plate is detected as a high temperature signal, while down-welling fluid from the cold top plate gives a low temperature signal.
This principle is employed here as well, however at a very fine resolution.
Figure~\ref{fig:Tmid_angle}(b) shows a colour plot of the temperature time series taken simultaneously by 11 thermocouples which are arranged in a semicircle at the side-wall at half height (see figure~\ref{fig:setup_cell}(c)).
This arrangement allows us to present a space-time-plot of the temperature with details never obtained before in a liquid metal flow experiment.
Temperatures below the average fluid temperature $\bar T$ are coloured in blue and temperatures above $\bar T$ are shown in orange.
The black dashed line re-plots the LSC orientation $\theta_\mathrm{LSC}=(\theta_\mathrm{top} + 180\degree + \theta_\mathrm{bot})/2$ from figure~\ref{fig:Tmid_angle}(a).
The profile has been additionally smoothed over time using a moving average filter over 5~successive measurements.
It confirms the coherence of the average LSC: Temperature at half height and velocity dynamics at top/bottom are in perfect synchronization and drift slowly as a common, single LSC roll.

Figures~\ref{fig:Tmid_angle}(c) and~(e) show magnified sections of panel~(b) at finer temporal resolution that correspond to those in panels~(f) and~(h), respectively.
In both cases, hot rising and cold falling plumes at the side-wall bounce together and move away from each other again.
This periodic motion is known as a sloshing mode of the LSC \citep{Zhou2009,Brown2009}.
We detect $\tau_\mathrm{slosh} = \tau_\mathrm{osc}$ from these two pairs of panels.
Previous experiments in water \citep{Zhou2009,Brown2009} report that the up- and down-welling flows come as close as 45\degree{}.
In our measurements this minimal azimuthal distance is much smaller, regularly reaching the order of our azimuthal resolution of 18\degree{}.
This extreme sloshing amplitude seems to be a property of low-$\Pran$ convection and the high inertia of liquid metals.

%----------------------------------------------------------------
\begin{figure}
\centerline{\includegraphics{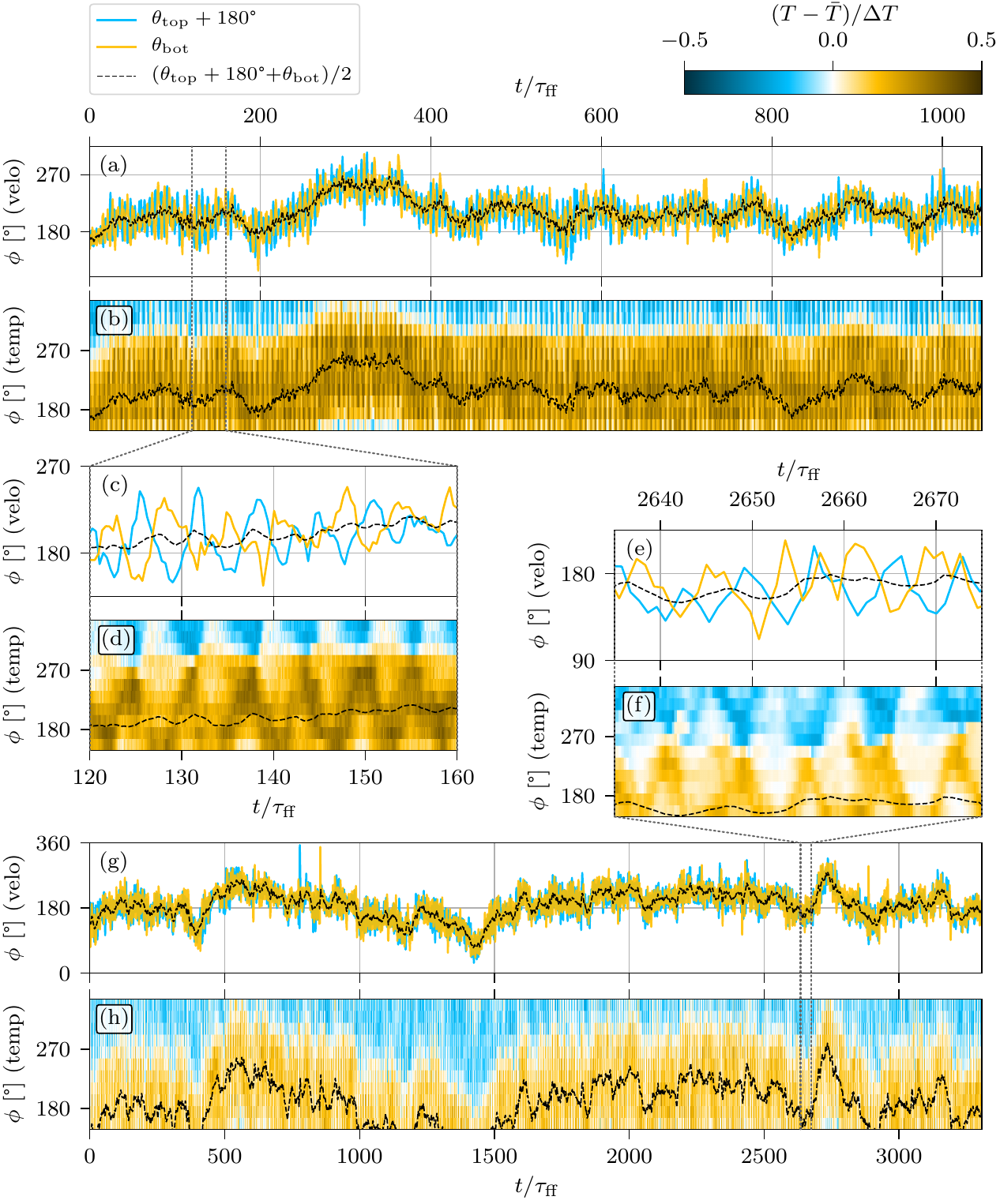}}
\caption{%
  Large-scale flow for $\Ray = 10^6$ ($\tau_\mathrm{ff} = 17$\,s, (a--d)) and $\Ray=10^7$ ($\tau_\mathrm{ff} = 5.4$\,s, (e--h)).
  Quantities and the colour scale are the same as in figure~\ref{fig:Tmid_angle}.
  (c) and (d) are detailed views of (a) and (b), respectively.
  (e) and (f) are detailed views of (g) and (h), respectively.
  The timespan of the detailed views is $40\tau_\mathrm{ff}$.}
\label{fig:Tmid_angle_lowmid}
\end{figure}
%----------------------------------------------------------------
A rare event in the large-scale flow dynamics is shown in figure~\ref{fig:Tmid_angle}(d) and~(g).
Here, the LSC rapidly changes its orientation by about 90\degree{} within less than $10\tau_\mathrm{ff}$ ($940 < t/\tau_\mathrm{ff} < 950$).
At the same time, the characteristic sloshing pattern in the temperature plot is disrupted and over most of the circumference the mean value of the temperature is detected which is in line with the absence of the coherent pattern up- or down-welling flow.
Only after the sudden orientation change the sloshing pattern reappears, now shifted by 90\degree\ in azimuthal direction.
This observation suggests that a cessation has taken place -- an event which has also been observed in experiments in water \citep{Brown2006} and fluorinert FC-77 electronic liquid \citep{Xie2013}.
Cessations consist of a breakdown of the coherent LSC into an incoherent flow state and a subsequent re-establishment of the coherence of the LSC with a different orientation.
These events are rare; in water experiments they occurred at rates of the order of days$^{-1}$ \citep{Brown2006, Xie2013}.
Our measurement series did not reach such time durations.
Consequently, a statistical analysis of cessations in the present liquid metal convection experiments was not possible.

In figure~\ref{fig:Tmid_angle_lowmid}, two additional experiments -- one at the lowest Rayleigh number~$Ra=10^6$ and one at an intermediate $Ra=10^7$ -- are presented.
The time series cover again more than 1000 free-fall times in both cases.
Just as for $Ra = 6\times 10^7$ in figure~\ref{fig:Tmid_angle}, the temperature and velocity measurements show the presence of a LSC, as well as sloshing and torsion modes in the magnifications.
Differences can mainly be seen with respect to the temperature magnitude when comparing the data to figure \ref{fig:Tmid_angle}.
With increasing $\Ray$, the amplitude of the fluid temperature at mid-height of the cell decreases steadily indicating an enhanced mixing of the scalar field due to an increasingly inertial fluid turbulence.
Furthermore, the hot and cold patches of the up- and down-wellings are thicker and more washed out for lower Rayleigh numbers.
This can be seen best for the lowest $\Ray = 10^6$ in figure~\ref{fig:Tmid_angle_lowmid}(d).
We have found that the magnitude of the total long-term azimuthal drift does not show a dependence on the Rayleigh number in the accessible range.
To summarize this part, our experimental study shows clearly that the LSC dynamics have to be considered as superposition of multiple processes with different characteristic time scales which can be reconstructed from the time series.

\subsection{Rayleigh number dependence of oscillation frequencies}
\label{sec:LSC_oscfreq}

\begin{figure}
  \centerline{\includegraphics{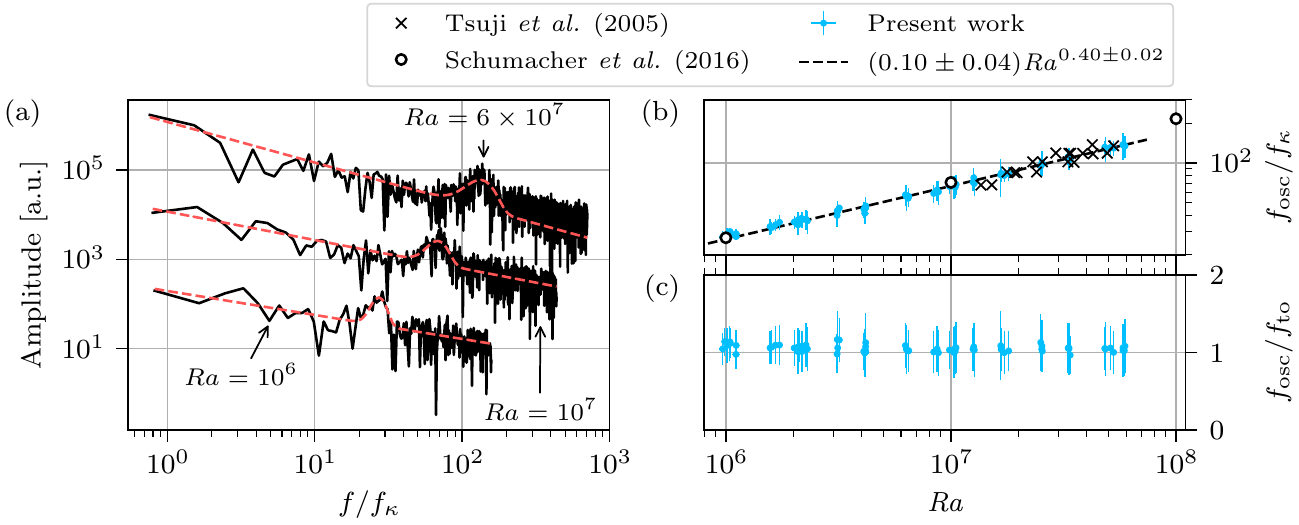}}
\caption{%
  (Colour online)
  (a)~Frequency spectra of $\theta_\mathrm{top}$ for different $\Ray$, normalized by the thermal diffusion frequency~$f_\kappa$.
  The spectra at $\Ray=10^6$ and $\Ray=6\times 10^7$ are shifted for a better visibility by a factor of $0.1$ and $20$, respectively.
  The dashed lines are fits of~\eqref{eq:charFreqFit} to the spectra to determine the characteristic oscillation frequency $f_\mathrm{osc}$.
  (b)~Scaling of the frequency $f_\mathrm{osc}$ with $\Ray$.
  The error bars are taken from the standard deviation~$\Delta f$ of the Gaussian fit~\eqref{eq:charFreqFit}.
  (c)~Comparison of $f_\mathrm{osc}$ to the turnover frequency~$f_\mathrm{to}$ over~$\Ray$.}
\label{fig:anglefreq(Ra)}
\end{figure}

Frequency spectra of $\theta_\mathrm{top}$ are shown for three Rayleigh numbers $\Ray$ in figure~\ref{fig:anglefreq(Ra)}(a).
A clear peak gives the oscillation frequency $f_\mathrm{osc}$ of the torsional mode.
The characteristic frequency value is extracted by fitting the function $A(f)$ to the spectra
%----------------------------------------------------------------
\begin{equation}
\label{eq:charFreqFit}
A(f) = a \exp\left(-\frac{(f-f_\mathrm{osc})^2}{2\Delta f^2}\right) +
  b\,f^\gamma \,.
\end{equation}
%----------------------------------------------------------------
This function models the spectra as a Gaussian peak at~$f_\mathrm{osc}$ with a width of~$\Delta f$ on top of an  algebraic power law background.
The fitting parameters are $a$, $b$, $\gamma$, $f_\mathrm{osc}$ and $\Delta f$.

Figure~\ref{fig:anglefreq(Ra)}(b) shows the $\Ray$-dependence of the frequency $f_\mathrm{osc}$, normalized by the thermal diffusion frequency
%----------------------------------------------------------------
\begin{equation}
f_\kappa = \frac{\kappa}{H^2} \,.
\end{equation} 
%----------------------------------------------------------------
The error bars correspond to the standard deviation~$\Delta f$ of the fitted Gaussian peaks.
A power law is fitted to the data using orthogonal distance regression~\citep[ODR,][]{Boggs1990} to incorporate the errors of both, the abscissa and the ordinate.
The error estimation of the fit is outlined in appendix~\ref{apx:PowerLawError}.
This procedure is used for all following power law fits in this work if not noted otherwise.
The fit results in a scaling of $f_\mathrm{osc}/f_\kappa \simeq (0.10 \pm 0.04) Ra^{0.40 \pm 0.02}$, indicated as a dashed line in figure~\ref{fig:anglefreq(Ra)}(b).
Dimensional arguments suggest a scaling of the oscillation frequency with the free-fall time~$\tau_\mathrm{ff}$ of $f_\mathrm{osc}\propto\Ray^{0.5}$ for constant material properties, since $1/\tau_\mathrm{ff} = \sqrt{\nu\kappa\Ray} /H^2$.
The exponent of 0.4 indicates, that the underlying time-scale is indeed close to the free-fall time, but that the inertial character of the fluid turbulence in the low-Prandtl-number flow affects the turbulent momentum transfer.
We will return to this point in section 4.2.

DNS by \citet{Schumacher2016} at $\Pran = 0.021$ result in a slightly stronger scaling of $f_\mathrm{osc}/f_\kappa \simeq (0.08 \pm 0.05) Ra^{0.42 \pm 0.02}$ when their data are corrected from radians to units of cycles per diffusive time (open circles in figure~\ref{fig:anglefreq(Ra)}(b)).
Previous experiments by \citet{Tsuji2005} in mercury at a Prandtl number of $\Pran = 0.024$ coincide with our results (crosses in figure~\ref{fig:anglefreq(Ra)}(b)).
For liquid gallium ($\Pran = 0.027$) in a $\Gamma = 2$ cell \citet{Vogt2018a} found the same scaling exponent as for the $\Gamma = 1$ case, but with a lower magnitude $f_\mathrm{osc}/f_\kappa \simeq 0.027 Ra^{0.419 \pm 0.006}$.
The scaling exponents and absolute values for measurements in larger Prandtl number fluids are generally higher.
For example, in water at $\Pran \sim 5.4$ scaling laws of $f_\mathrm{osc}/f_\kappa \simeq 0.2 \Ray^{0.46}$ \citep{Qiu2001}, $f_\mathrm{osc}/f_\kappa \simeq 0.167 Ra^{0.47}$ \citep{Qiu2004}, and $f_\mathrm{osc}/f_\kappa \simeq 0.12 \Ray^{0.49}$ \citep{Zhou2009} are found.
Experiments in methanol \citep[$Pr=6.0$]{Funfschilling2004} give a similar result of $f_\mathrm{osc}/f_\kappa \simeq 0.126 \Ray^{0.460\pm0.012}$.
Considering that the thermal diffusivity of water~\citep{Cengel2008} is about 70 times lower than that of GaInSn at 35\,\degree{}C, the absolute frequencies $f_\mathrm{osc}$ in water at $\Ray=10^7$ are about 14 times smaller than our values in GaInSn.
This higher intensity of the flow dynamics in liquid metals is to be expected, as low-$\Pran$ liquids are known to drive a more turbulent convective flow than high-$\Pran$ fluids at the same $\Ray$~\citep{Breuer2004,Schumacher2015}.

The averaged velocity components (\ref{eq:LSCvelocomp}) are also used to calculate the velocity amplitude of the LSC 
%----------------------------------------------------------------
\begin{equation}
v_\mathrm{LSC} 
  = \left\langle\frac{|\vec v_\mathrm{top}| + |\vec v_\mathrm{bot}|}
                     {2}\right\rangle_t \,,
\end{equation} 
%----------------------------------------------------------------
where $\langle \cdot \rangle_t$ denotes a time average.
Using this velocity, the turnover time~$\tau_\mathrm{to}$ of the LSC can be defined as $\tau_\mathrm{to} = \pi H/v_\mathrm{LSC}$.
Here, a roll shape in form of a circle of diameter~$H$ has been assumed as the LSC path.
The turnover frequency is then 
%----------------------------------------------------------------
\begin{equation}
f_\mathrm{to} = \frac{v_\mathrm{LSC}}{\pi H} \,.
\end{equation} 
%----------------------------------------------------------------
Figure~\ref{fig:anglefreq(Ra)}(c) shows the ratio $f_\mathrm{osc}/f_\mathrm{to}$, which is close to unity for all Rayleigh numbers.
This implies that one period of the torsional mode -- and thus also of the sloshing mode -- takes one turnover time of the LSC.
If, alternatively, the length $2H+2D$ is used for the LSC path instead of the circumference of a circle, $f_\mathrm{to}$ decreases by a factor of $\pi/4 \approx 0.79$.
However, the relation of $f_\mathrm{osc} \propto f_\mathrm{to}$ would still be valid.

\subsection{Interplay of the torsion and sloshing modes}
\label{sec:LSC_modeinterplay}

%-------------------------------------------------------
\begin{figure}
\centerline{\includegraphics{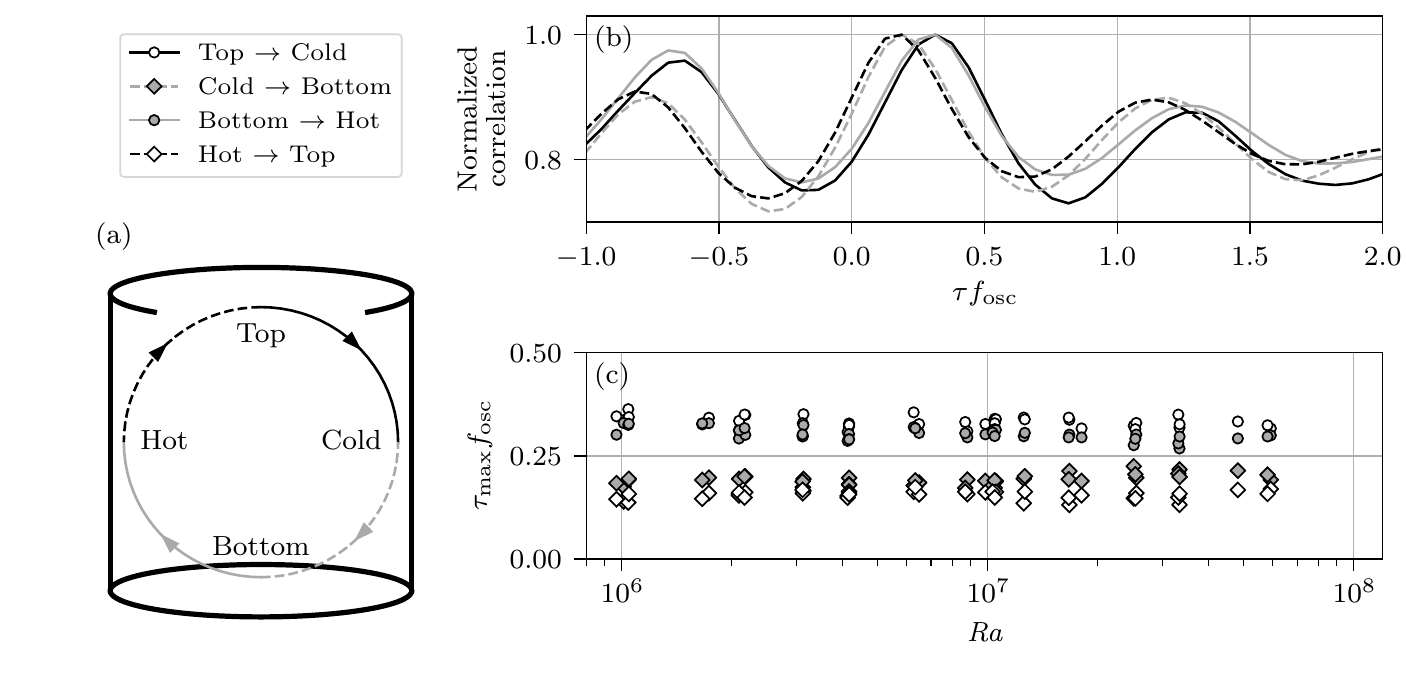}}
\caption{%
  Correlation of temperatures at mid height with flow orientations at the top and bottom plates.
  (a)~Illustration of the possible correlations.
  (b)~Example of correlations which are normalized by their maxima and plotted over the time shift~$\tau$ in units of the oscillation frequency~$f_\mathrm{osc}$.
  The data record is taken from the measurement in figure~\ref{fig:Tmid_angle} at $\Ray=6\times 10^7$.
  (c)~Time shift~$\tau_\mathrm{max}$ of the maximal correlation over $\Ray$.}
\label{fig:mode_progression}
\end{figure}
%------------------------------------------------------- 

In this section, we investigate how the sloshing and torsion modes described in section~\ref{sec:LSC_structure} coexist and build a single coherent flow structure.
The basic connection of the two modes is that the flow directions at the top and bottom plate, $\theta_\mathrm{top}$ and $\theta_\mathrm{bot}$, indicate the azimuthal position where the up- and down-welling flows will appear. This can be monitored by the temperature sensors at mid-height.
We have already shown, that on average the top and bottom flows are anti-parallel with $\langle\theta_\mathrm{bot}\rangle_t - \langle\theta_\mathrm{top}\rangle_t \sim 180$\degree{} (see figure~\ref{fig:Tmid_angle}(a)).
If one assumes that these flows will be deflected in different azimuthal directions by an angle $\Delta\theta$ (which stands for the effect of the LSC torsion) then the orientation angles will get closer to one another.
With $\theta_\mathrm{top/bot} = \langle\theta_\mathrm{top/bot}\rangle_t \pm \Delta\theta$, the azimuthal distance of the up- and down-welling flows is given by
%-------------------------------------------------------
\begin{equation}
\theta_\mathrm{bot} - \theta_\mathrm{top} \sim 180\degree{} - 2 \Delta\theta \,.
\end{equation}
%-------------------------------------------------------
With the torsion displacement $\Delta\theta$ at top and bottom getting closer to 90\degree{}, the hot up- and cold down-welling flows will also get closer to each other.
This is exactly the behaviour of the sloshing mode detected in figure~\ref{fig:Tmid_angle}(b).

However, the flow orientation at the plates and the respective vertical flow at half-height of the cell do not coincide in time.
We have established, that one oscillation period is the same as one turnover of the LSC.
From this point, we would expect that the flow orientation propagates with the same speed, e.g., a given flow direction at the top plate would result in the same azimuthal position of the down-flow a quarter turnover later.
To investigate this behaviour more closely, we calculate a temporal correlation of the top and bottom LSC angles with the temperatures measured at mid height (see figure~\ref{fig:mode_progression}(a)).
As an example, we correlate $\theta_\mathrm{top}$ with the cold down-flow at mid height, which we denote as Top $\to$ Cold.
All positive values of the temperature profile are clipped in figure~\ref{fig:Tmid_angle}(b) to zero, in order to include the cold signature of the down-flow only.
This is equivalent to clipping the temperatures to the mean fluid temperature $\bar T$.
Next, we construct a pseudo-temperature profile $\mathcal{T}$ from the time series of $\theta_\mathrm{top}$ which is given by
%-------------------------------------------------------
\begin{equation}
\mathcal{T}(\phi, t) = 
  \begin{cases}
  A \cos^2\bigl(\phi-\theta_\mathrm{top}(t)\bigr) & \text{for }
    \theta_\mathrm{top}-90\degree{} < \phi < \theta_\mathrm{top}+90\degree \\
  0 & \text{else}
  \end{cases} \,.
\end{equation}
%-------------------------------------------------------
This profile emulates the temperature at mid height if the position of the down-flow would follow the flow orientation at the top plate instantaneously.
The amplitude is set to $A=-1$ in this particular case since we correlate with the negative values of the measured temperature profile.
In case of a correlation with the up-welling flow, an amplitude $A=1$ is taken.
The function $\mathcal T$ is evaluated at the azimuthal positions of the mid-height thermocouples and correlated with the clipped temperature profile over time.
The result is normalized by its maximal value and plotted in figure~\ref{fig:mode_progression}(b) as a solid black line.
The correlation time shift $\tau$ is normalized by the oscillation frequency $f_\mathrm{osc}$.

We observe that the maximum correlation is shifted to $\tau f_\mathrm{osc}>0$, i.e., the down-flow at mid height lags behind the flow orientation at the top.
The time lag~$\tau_\mathrm{max}$ of the maximum correlation is calculated by fitting a quadratic polynomial around the maximal value.
In the case Top $\to$ Cold the time shift is $\tau_\mathrm{max}f_\mathrm{osc}  = 0.32$, which is larger than the expected value of $1/4$.
The correlations of the three other cases Cold $\to$ Bottom, Bottom $\to$ Hot and Hot $\to$ Top are displayed in figure~\ref{fig:mode_progression}(b) as well.
It can be seen that the vertical flows at mid height have a time lag towards the flows at the respective plate, which is larger than a quarter of one oscillation period (Top $\to$ Cold and Bottom $\to$ Hot).
Furthermore, the horizontal flows at the plates follow the mid-height flows with a lag shorter than a quarter oscillation period.
We repeated this analysis for all our experiments.
The corresponding time shifts display this behaviour (see figure~\ref{fig:mode_progression}(c)) in all data sets.
On average the time lags~$\tau_\mathrm{max}f_\mathrm{osc}$ are: $0.33\pm0.01$ for Top $\to$ Cold, $0.19\pm0.01$ for Cold $\to$ Bottom, $0.30\pm0.01$ for Bottom $\to$ Hot, and $0.16\pm0.01$ for Hot $\to$ Top.
The sum of all four time lags gives on average the expected value of one, here $0.98 \pm 0.03$, and thus the propagation of the flow orientation requires the same amount of time as one oscillation period or one turnover.
From the present data, the flow can be understood as independent parcels of the fluid circulating in the cell.
Each fluid parcel circulates on average in a vertical plane which it does not leave.
The observed flow modes are then a result of the collective motion of these fluid parcels, which are phase shifted in time and the azimuthal orientation of their plane.
This interpretation of our data does not predict significant azimuthal velocity components as part of the mode dynamics of the LSC.
This cannot be verified with the current set-up, but future experiments could aim to include the measurement of azimuthal velocities.

A similar analysis as in figure~\ref{fig:mode_progression} is found in \cite{Qiu2004}.
Their data reveal a shift close to $\tau_\mathrm{osc}/4$ between fluctuations of the velocity at the bottom centre and of the temperature at half-height near the up-welling flow. 
This would correspond to our Bottom $\to$ Hot case. 
An exact shift value is however not discernable from their plots. 
Other data on the phase shift between torsion and sloshing modes can be grouped into three categories:
(i)~Correlation of the top and bottom flow orientations, which gives the characteristic torsion phase shift of $\tau_\mathrm{osc}/2$.
See experiments by~\cite{Funfschilling2004} in methanol at $\Pran = 6.0$, by~\cite{Xie2013} in fluorinert FC-77 electronic liquid at $\Pran = 19.4$, or by~\cite{Khalilov2018} in liquid sodium at $\Pran = 0.0094$.
(ii)~Correlation of the hot and cold temperature signatures at half height, resulting in the characteristic sloshing phase shift of $\tau_\mathrm{osc}/2$.
See data by \cite{Qiu2001} and \cite{Xi2009} in water at $\Pran \sim 5.3$.
(iii)~Correlation of the mean LSC orientation at half height and the top or bottom flow orientation yielding a phase shift of $\tau_\mathrm{osc}/4$.
See measurements by \cite{Zhou2009} in water at $\Pran = 5.3$.  
These results are consistent with our data:
(i)~and (ii) are equivalent to the sum of two successive phase shifts in figure~6 (e.g. Top $\to$ Cold plus Cold $\to$ Bottom).
(iii) is equal to an average of two successive phase shifts in figure~6.
Each of these analysis steps averages out the deviations of individual time lags from $\tau_\mathrm{osc}/4$ and result in phase shifts of $\tau_\mathrm{osc}/2$ for (i)~and (ii), and $\tau_\mathrm{osc}/4$ for (iii).
All these data show that the torsion and sloshing modes are present for a wide range of Prandtl numbers.
However, whether the specific time lags found in our system are a general feature needs to be clarified in the future.

%-------------------------------------------------------
\begin{figure}
\centerline{\includegraphics{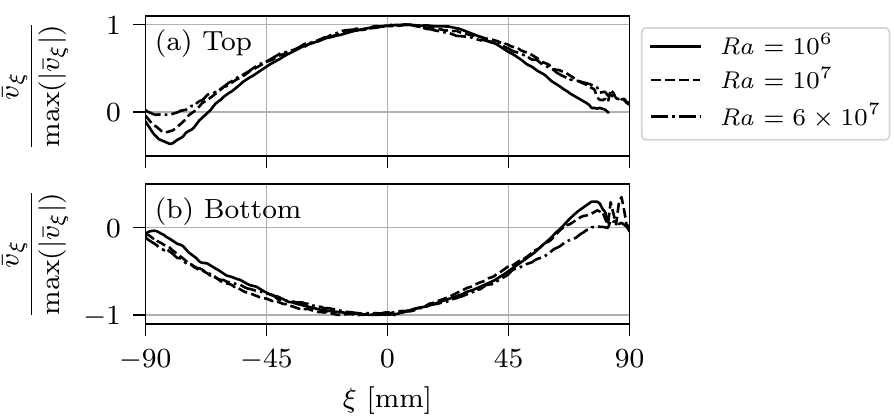}}
\caption{%
  Averaged and normalized horizontal UDV velocity profiles along the LSC at (a) the top and (b) the bottom plate for different $\Ray$.
  The experiments correspond to the ones shown in figure~\ref{fig:anglefreq(Ra)}(a).}
\label{fig:LSCprofile}
\end{figure}
%------------------------------------------------------- 
The deviations of the individual lags from a quarter oscillation period indicate that the LSC path is asymmetric.
Such distortions might be caused by corner vortices of the LSC which are observed in simulations.
Here, we introduce a horizontal axis~$\xi$ that is aligned in direction of the LSC orientation.
To calculate an averaged horizontal velocity profile of the LSC at the top (bottom) plate, we select the velocity profiles of the UDV sensors T\textsubscript{0}, T\textsubscript{45}, and T\textsubscript{90} (B\textsubscript{0}, B\textsubscript{45}, and B\textsubscript{90}) at time instants when the LSC orientation $\theta_\mathrm{top}$ ($\theta_\mathrm{bot}$) is aligned within $\pm 5\degree$ with the azimuthal position of a sensor.
The averages of those velocity profiles $\bar v_\xi(\xi)$ are normalized by their maximum magnitude and shown in figure~\ref{fig:LSCprofile} for three different $\Ray$.
It should be noted, that the noisy or missing profiles for $\xi > 80$\,mm are due to an inaccessible zone close to the UDV-sensors caused by the ringing of the piezo-crystal in the transducer.
The velocities at the top plate are predominantly positive (flow to the right) and the bottom velocities are negative (flow to the left).
Inverted velocities can only be seen at the positions where the up- and down-welling flows are impinging on the plates (left at the top, right at the bottom).
These profiles indicate indeed the presence of recirculation vortices in the cell corners, which seem to become smaller with increasing $\Ray$.
This challenges our supposition at the beginning of this paragraph that corner vortices might be responsible for the different time lags in figure~\ref{fig:mode_progression}(c).
With increasing Rayleigh number the fluid turbulence becomes more vigorous.
This means that the vortical structures are still there, but get averaged out more effectively.

A further explanation for the phase shift deviations from the expected value of $\tau_\mathrm{osc}/4$ could be traced to a varying LSC velocity.
The fluid would have to be slower when leaving the plates (Top $\to$ Cold and Bottom $\to$ Hot) and accelerate while approaching the opposite plate (Cold $\to$ Bottom and Hot $\to$ Top).
This point has to be left for future studies on this subject.

\section{Turbulent transport of momentum and heat}
\label{sec:transport}

\subsection{Heat transport}
\label{sec:transport_heat}

\begin{figure}
\centerline{\includegraphics{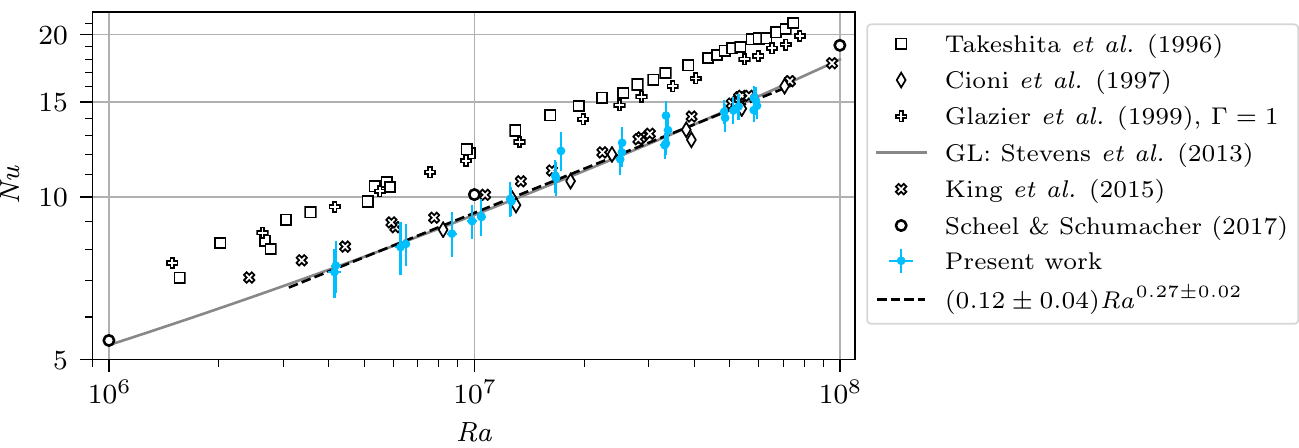}}
\caption{%
  (Colour online)
  Turbulent heat transfer in the liquid metal convection flow.
  The scaling of $\Nuss$ vs. $\Ray$ is displayed.
  $\Nuss$ is determined from the temperature difference of the cooling water for $T_\mathrm{out}-T_\mathrm{in} > 0.2$\,K.
  Our data are compared with a numerical simulation~(circles), four laboratory experiments and the GL theory at $\Pran=0.029$, as indicated in the legend.}
\label{fig:Nu(Ra)}
\end{figure}

The present section discusses the global transport properties in the liquid metal convection flow and compares the results with other experiments and simulations.
We first analyse the turbulent heat transport in the experiment.
The heat flux through the fluid layer is characterized by the Nusselt number~$\Nuss$ which is calculated at the cooled plate, $\Nuss = \dot Q/\dot Q_\mathrm{cond}$.
Here, $\dot Q_\mathrm{cond} = \lambda \pi R^2\Delta T/H$ is the purely conductive heat flux.
The data are plotted over $\Ray$ in figure~\ref{fig:Nu(Ra)} and we find $\Nuss \simeq (0.12 \pm 0.04) \Ray^{0.27 \pm 0.02}$.
It agrees excellently with measurements by \citet{Cioni1997} in mercury ($\Pran=0.025$) and matches the numerical results by \citet{Scheel2017} of $\Nuss \simeq (0.13 \pm 0.04) \Ray^{0.27 \pm 0.01}$ for $\Pran=0.021$ with only a small shift.
%----------------------------------------------------
%-----------------------------------------------------

The results by \cite{Takeshita1996} in mercury ($\Pran=0.024$) agree with respect to the scaling exponent, but give a somewhat higher Nusselt number magnitude with $\Nuss \simeq 0.155 \Ray^{0.27\pm0.02}$. 
\cite{King2013,King2015} found the same magnitude of $\Nuss$ in gallium ($\Pran\sim0.025$), but their scaling exponent is smaller, $\Nuss \simeq (0.19\pm0.01)\Ray^{0.249\pm0.004}$.
All of the above experiments and simulations were conducted in cylindrical cells with $\Gamma = 1$.
Finally, \cite{Glazier1999} conducted experiments in mercury ($\Pran \sim 0.025$) for multiple aspect ratios.
Their data at $\Gamma = 1$ are closer to the results by \cite{Takeshita1996} in magnitude (see figure~\ref{fig:Nu(Ra)}). 
A least-squares fit gives a slower scaling $\Nuss \propto \Ray^{0.242\pm0.002}$ in the considered range. 
Including all their data for $\Gamma=0.5$ and~$2$, they found a scaling $\Nuss \propto \Ray^{0.29\pm0.01}$ over a large range of $10^5 < \Ray < 10^{11}$. 
The deviation of their $\Gamma=1$ data from this exponent is attributed to a strong bulk circulation.

The Grossmann-Lohse theory (GL) prediction \citep{Stevens2013} is in very good agreement with our data as seen by the grey line in figure~\ref{fig:Nu(Ra)}. 
It can also be seen, that GL predicts an increase in the exponent of $\Nuss(\Ray)$ for higher $\Ray$.
At smaller Rayleigh numbers ($\Ray\lesssim 5\times 10^5$), \cite{Rossby1969} found $\Nuss \simeq 0.147 \Ray^{0.257\pm0.004}$ for $\Pran=0.025$ and $\Gamma\sim22$ and $7.4$. 
\cite{Kek1993} reported $\Nuss \simeq 0.20\Ray^{0.20}$ at $Pr=0.006$ and $\Gamma \sim 11$.

\subsection{Momentum transport}
\label{sec:transport_momentum}
The turbulent momentum transport of the convection flow is quantified by the Reynolds number~$\Rey$.
In contrast to the quantification of the turbulent heat transfer by a Nusselt number, the Reynolds number is not uniquely defined since different characteristic velocities can be employed for its definition.
With the multiple probes at hand, we can define three different $\Rey$ with three different corresponding characteristic velocities:
1) the typical horizontal velocity magnitude near the plates, 2) the typical vertical velocity magnitude of the LSC along the side wall and 3) the turbulent velocity fluctuations in the centre of the cell,
%-------------------------------------------------------
\begin{equation}
\Rey_\mathrm{LSC} = \frac{ v_\mathrm{LSC}H}{\nu} \,, \quad 
\Rey_\mathrm{vert} = \frac{ v_\mathrm{vert}H}{\nu} \,, \quad \mbox{and} \quad 
\Rey_\mathrm{centre} = \frac{ v_\mathrm{centre}H}{\nu} \,.
\end{equation}
%-------------------------------------------------------
This circumstance opens the opportunity to directly compare the sensitivity of the scaling exponent with respect to these different characteristic velocities.

\begin{figure}
\centerline{\includegraphics{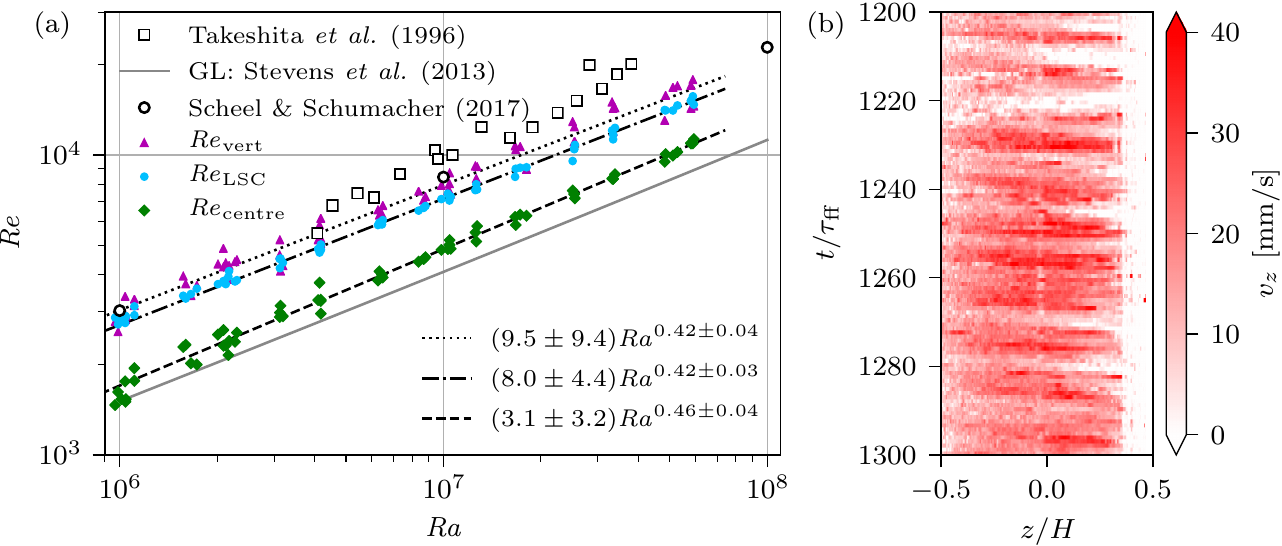}}
\caption{%
  (Colour online)
  Analysis of the turbulent momentum transfer in the liquid metal convection flow based on the present UDV measurements.
  (a)~Scaling of Reynolds number~$\Rey$ with Rayleigh number~$\Ray$.
  Uncertainties of the present measurements were removed for better visibility.
  Their relative values are about 20 to 30\,\% due to the turbulent fluctuations.
  (b)~Colour-plot of $v_z$ measured by UDV sensor V\textsubscript{0} for $Ra = 6 \times 10^7$.
  The data correspond to the measurement shown in figure~\ref{fig:Tmid_angle}.}
\label{fig:Re(Ra)}
\end{figure}

First, the large scale flow is characterized by the velocity magnitude $v_\mathrm{LSC}$ as done in  section~\ref{sec:LSC_oscfreq}.
The resulting Reynolds number $\Rey_\mathrm{LSC} = v_\mathrm{LSC}H/\nu$ is plotted in figure~\ref{fig:Re(Ra)} (filled circles).
A power law fit reveals a scaling of $\Rey_\mathrm{LSC}\simeq (8.0\pm 4.4) Ra^{0.42\pm 0.03}$.
The exponent is close to the scaling of $f_\mathrm{to}$ and $f_\mathrm{osc}$ in figure~\ref{fig:anglefreq(Ra)}(b) since $\Rey_\mathrm{LSC} = f_\mathrm{to} \pi H^2/\nu$.

Second, the vertical velocity of the LSC is measured at radial position $r/R = 0.8$ by the UDV sensor V\textsubscript{0}.
Due to the sloshing mode, the up- or down-welling flows move periodically towards and away from the sensor measuring volume (see figure~\ref{fig:Re(Ra)}(b)).
Additionally, the LSC is slowly rotating as a whole.
Since a pronounced vertical flow of the LSC is of interest only, we estimate the characteristic vertical velocity by calculating the average velocity profile $v_z(t)$ over an interval $H/4$ centred around the mid-plane for every time step, similar to (\ref{eq:LSCvelocomp}).
The corresponding standard deviation ($\mathrm{std}$) is added to the average of the resulting velocity magnitude to accommodate for the fluctuations of this signal.
Thus, $v_\mathrm{vert} = \langle |v_z(t)| \rangle_t+\mathrm{std}(|v_z(t)|)$.
This velocity gives a vertical Reynolds number $\Rey_\mathrm{vert} = v_\mathrm{vert}H/\nu$.
The values are shown in figure~\ref{fig:Re(Ra)}(a) as triangles and give a scaling of $\Rey_\mathrm{vert}\simeq (9.5 \pm 9.4) Ra^{0.42 \pm 0.04}$.
%-------------------------------------------------------
%------------------------------------------------------- 

Finally, the turbulent velocity fluctuations are considered in the centre of the cell.
Here, the three crossing UDV sensors M\textsubscript{0}, M\textsubscript{90}, and V\textsubscript{c} can measure all three components of the velocity vector $\vec v_\mathrm{centre}$ (see figure~\ref{fig:setup_cell}(b)).
The components are determined by taking the root-mean-square (rms) value of, again, the central $H/4$ interval of the velocity profiles.
The rms time-average of the velocity magnitude $v_\mathrm{centre} = \langle(\vec v_\mathrm{centre}(t))^2\rangle^{1/2}_{t}$ is used to calculate $\Rey_\mathrm{centre} = v_\mathrm{centre}H/\nu$.
It is plotted as diamonds in figure~\ref{fig:Re(Ra)}(a), along with the power-law fit of $\Rey_\mathrm{centre} \simeq (3.1\pm 3.2) Ra^{0.46\pm 0.04}$.

The scaling exponent of $\Rey_\mathrm{centre}(\Ray)$ agrees very well with the result of DNS by \citet{Scheel2017} (open circles in figure~\ref{fig:Re(Ra)}(a)).
In the numerical simulations, the Reynolds number was calculated from the rms-velocity over the whole cell for $\Pran = 0.021$ and gave a scaling of $\Rey\simeq (6.5 \pm 0.6) Ra^{0.45 \pm 0.01}$.
The absolute values of $\Rey_\mathrm{centre}$ are about half as large as the results of the DNS, since the average over the whole cell volume also includes the high-velocity components of the LSC outside the centre region.
The absolute values of $\Rey_\mathrm{LSC}$ and $\Rey_\mathrm{vert}$ match the DNS results more closely, but have a somewhat smaller exponent of $0.42$.

How do our results compare to previous laboratory experiments? 
Measurements of a vertical Reynolds number in mercury by \citet{Takeshita1996} were also taken at half height and $r/R=0.8$.
They show a higher velocity magnitude with a scaling of $\Rey\simeq 6.24 Ra^{0.46 \pm 0.02}$.
In the $\Gamma = 2$ case, \citet{Vogt2018a} found an increased scaling exponent for a horizontal $\Rey$ at the cell centre: $\Rey\simeq 5.662 Ra^{0.483}$.
Comparing all these results underlines the dependence of the momentum transport on the specific velocity that enters the Reynolds number definition.
The somewhat smaller scaling exponent for the horizontal and vertical LSC velocities in comparison to previous experiments or the DNS can thus be considered as a result of probing different parts of the complex three-dimensional flow structure that we analysed before, as well as using different measurement techniques and procedures of calculating the characteristic velocities.
Interestingly, the turbulent fluctuations seem to be the best indication of the global momentum transport scaling as reported by DNS, albeit being smaller in their absolute magnitude.

We also show the Reynolds number predicted by the GL theory~\citep{Stevens2013} in figure~\ref{fig:Re(Ra)}.
Compared to $\Rey_\mathrm{LSC}$ and $\Rey_\mathrm{vert}$, the GL theory underpredicts the Reynolds number by up to a factor of~2.
\cite{Stevens2013} used here a measurement at $Pr=5.4$ to fix the values of $\Rey$.
A least-squares power law fit of the GL results in figure~\ref{fig:Re(Ra)} gives an exponent of $\Rey\propto\Ray^{0.44}$, which lies within the accuracy of our experimental data.

\section{Small-scale flow structure in the centre of the convection cell}
\label{sec:result_turbulence}

%------------------------------------------------------- 
\begin{table}
\centering
\begin{tabular}{c p{10pt} ccc p{10pt} ccc p{10pt} ccc}
&& \multicolumn{3}{c}{$\Ray = 10^6$} &&
   \multicolumn{3}{c}{$\Ray = 10^7$} &&
   \multicolumn{3}{c}{$\Ray = 3.3\times 10^7$} \\
\hline
  $\tilde r = \dfrac{r}{R}$ &
  & $\dfrac{S_\xi(\tilde r)}{S_z(\tilde r)}$
  & $\dfrac{S_\eta(\tilde r)}{S_z(\tilde r)}$
  & $\dfrac{S_\eta(\tilde r)}{S_\xi(\tilde r)}$ &
  & $\dfrac{S_\xi(\tilde r)}{S_z(\tilde r)}$
  & $\dfrac{S_\eta(\tilde r)}{S_z(\tilde r)}$
  & $\dfrac{S_\eta(\tilde r)}{S_\xi(\tilde r)}$ &
  & $\dfrac{S_\xi(\tilde r)}{S_z(\tilde r)}$
  & $\dfrac{S_\eta(\tilde r)}{S_z(\tilde r)}$
  & $\dfrac{S_\eta(\tilde r)}{S_\xi(\tilde r)}$\\[15pt]
0.01 && 0.56 & 0.30 & 0.53 && 0.77 & 0.38 & 0.49 && 0.83 & 0.69 & 0.83 \\
0.10 && 0.98 & 0.77 & 0.79 && 1.15 & 0.82 & 0.71 && 1.13 & 0.93 & 0.83 \\
0.50 && 0.82 & 1.01 & 1.23 && 1.03 & 0.90 & 0.88 && 0.99 & 0.93 & 0.93 \\
\hline
\end{tabular}
\caption{%
  Comparison of second order structure functions (or second order velocity increment moment) evaluated at three distances in the centre of the convection cell for different directions.
  Ratios of the vertical structure function $S_z(\tilde r)$, the horizontal structure function parallel to the LSC $S_\xi(\tilde r)$, and the horizontal structure function perpendicular to the LSC  $S_\eta(\tilde r)$ are analysed.
  The table lists the ratios of the horizontal to the vertical values as well as the ratio of both horizontal function values.
  The data are given for three Rayleigh numbers~$\Ray$.
  The spatial distance~$r$ is given in units  of the convection cell radius~$R$ and $\tilde r=r/R$.}
\label{tab:structFrac}
\end{table}

The UDV measurements in liquid metal convection monitor the longitudinal velocity profile along the beam line.
This opens the possibility to analyse the statistics of longitudinal velocity increments directly and thus the small-scale statistics in the bulk of a liquid metal flow (see \cite{Lohse2010} for a comprehensive review).
We consider longitudinal velocity increments which are given by $\delta_{\vec r} v(t) = [\vec v(\vec x+\vec r,t)-\vec v(\vec x,t)]\cdot \vec r/r$ with $r=|\vec r|$.
The velocity profiles $v$ are measured by the UDV sensors M\textsubscript{0}, M\textsubscript{90}, and V\textsubscript{c}.
Vertical velocity profiles~$v_z(z,t)$ are taken directly from the UDV sensor V\textsubscript{c}.
Horizontal velocity profiles, however, have to be considered in their relation to the LSC orientation.
In line with the discussion in section~\ref{sec:LSC_modeinterplay}, we collect the velocity profiles~$v_\xi(\xi, t)$ of the UDV sensors M\textsubscript{0} and M\textsubscript{90} when the mean LSC orientation angle $\theta_\mathrm{LSC}$ is aligned within $\pm 5\degree$ with the beam line  of these sensors.
Additionally, we introduce a horizontal axis~$\eta$, which is perpendicular to the LSC orientation and collect the horizontal velocity profiles~$v_\eta(\eta, t)$ from M\textsubscript{0} and M\textsubscript{90} when $\theta_\mathrm{LSC}+90\degree$ is aligned with the sensors.
The analysis is thus conditioned to the LSC orientation.

In order to quantify the degree of isotropy of the velocity fluctuations in the turbulent flow, we start with a comparison of the longitudinal second order structure functions or velocity increment moments which are given by 
%-------------------------------------------------------
\begin{equation}
S_i (r) = \left\langle(\delta_{\vec r} v_i)^2\right\rangle_{\vec x,t} \quad\text{with}\quad \vec r = r \vec e_i  \,,
\end{equation}
%-------------------------------------------------------
and $i = \xi, \eta, z$.
The average is taken with respect to time and to points along the beam line.
Table~\ref{tab:structFrac} shows the mutual ratios of these moments for three distances ~$\tilde r=r/R$ with $R$ being the cell radius.
Approximate isotropy would follow when these ratios are very close to unity.
Our data clearly indicate that this is not the case, particularly for the smallest separation.
The data suggest that the LSC flow is responsible for these deviations and affects the fluctuations over a wide range of scales.
The present Rayleigh numbers are small in comparison to the measurements by \cite{Mashiko2004} at $\Pran=0.024$, by \cite{Sun2006} at $\Pran=4.3$, or the direct numerical simulations by \cite{Kunnen2008a} at $\Pran=4$ such that we cannot present a scaling analysis of the second order structure function.

\begin{figure}
\centerline{\includegraphics{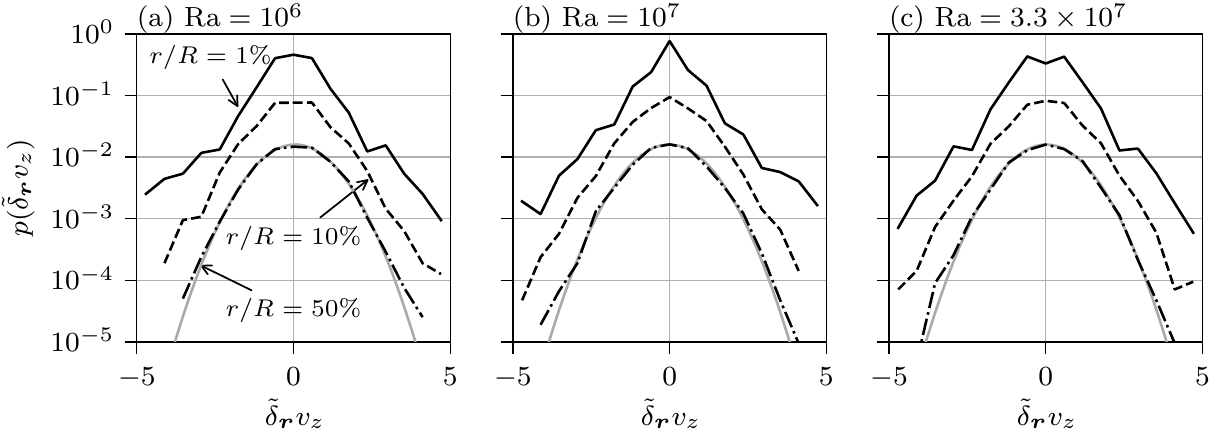}}
\caption{%
  Probability density functions (PDF) of the normalized velocity increments~$\tilde\delta_{\vec r} v_z$, calculated from the UDV sensor V\textsubscript{c}.
  The plots (a) to (c) show different Rayleigh numbers.
  The styles of the black lines correspond to the spatial increment~$r$ as specified in (a).
  The PDF for higher $r$ are successively shifted by a factor of 0.2 for a better visibility.
  The grey line is a normal distribution with mean and variance of the PDF for $r/R = 50\%$.}
\label{fig:veloincr}
\end{figure}

Figure~\ref{fig:veloincr} shows the probability density functions~(PDF) of the normalized vertical velocity increment $\tilde\delta_{\vec r} v_z = \delta_{\vec r} v_z/\langle\delta_{\vec r} v_z\rangle_\mathrm{rms}$ for the same separations~$\tilde r$ and Rayleigh numbers~$\Ray$ as in table~\ref{tab:structFrac}.
We have verified that the quantitative behaviour of the corresponding PDFs of $\tilde\delta_{\vec r} v_{\xi}$ and $\tilde\delta_{\vec r} v_{\eta}$ is the same (and thus not displayed). For all values of $\Ray$ the PDF approaches a normal distribution with increasing increment size, a result which is well-known from homogeneous isotropic turbulence (see e.g. the DNS by \cite{Gotoh2002}).
For the smallest separation the PDF is characterized by exponential tails.
Even though we were not able to resolve far tails of the PDFs, which are stretched exponential, our distributions reflect an intermittent, fully developed fluid turbulence in the bulk of the cell.
%------------------------------------------------------- 
%------------------------------------------------------- 

\section{Conclusion}
\label{sec:conclusion}

We presented an analysis of turbulent liquid metal convection in a cylindrical cell with aspect ratio $\Gamma = 1$.
The combination of multiple UDV and temperature measurements allowed us to reconstruct and characterize essential features of the 3D large-scale circulation in the cell experimentally.
The dynamics of the LSC is a superposition of different dynamical processes on different time scales.
The slow meandering of the large-scale flow orientation in the closed cell proceeds at a scale of the  order of up to $10^3$ free-fall time units in the present parameter range -- a time scale that is not accessible in DNS at such low Prandtl numbers.
The torsional and sloshing modes of the LSC, which are known from studies in water, could be detected by velocity and temperature measurements in the present low-$\Pran$ experiment.
We find a very synchronous sloshing at a time scale of $10 \tau_\mathrm{ff}$ that suggests a more coherent large-scale flow as for higher $\Pran$.
This time scale is consistent with temperature measurements of older experiments in mercury~\citep{Tsuji2005} and recent DNS~\citep{Schumacher2016}.
It also coincides with the turnover time of the LSC.
The temporal correlations between different segments of the LSC, which we verified by combination of velocity and temperature measurements are nearly independent of the Rayleigh number.
Cessations of the large-scale flow remain rare events in the low-Prandtl-number convection regime.
In summary, our analysis supports the picture of a very coherent large-scale flow in the closed cell for turbulent convection at such low Prandtl number.

The turbulent momentum transport was determined in multiple ways in the cell by {\em direct} velocity measurements.
Depending on the UDV beam line position with respect to the large-scale flow, the resulting $\Rey$ can vary by a factor of two or even more in amplitude.
The scaling behaviour with respect to the Rayleigh number is found to agree well with previous studies.
The same holds for the turbulent heat transfer.
The present ultrasound measurements of velocity increments reveal an inertial fluid turbulence in the bulk, e.g., by extended tails of the distribution of small-separation increments.
We also showed that coherent large-scale flow seems to prevent the establishment of local isotropy in the bulk, a point that might be worth to be deeper explored in simulations, in particular for larger aspect ratios.

Our study is based to a large part on direct velocity field measurements in the liquid metal flow.
Even though several of the properties that we discussed are known from convective turbulence in air or water, one value of this work is to demonstrate these features in an opaque liquid metal flow.
Differences of the presented data with respect to data taken in water or air are: (i) a more coherent dynamics of the LSC, and (ii) large-scale flow oscillations with higher frequencies and amplitudes, most probably due to the higher inertia of fluid turbulence in liquid metal flows.
The latter point is also apparent in the higher amplitudes of the mean global momentum transfer.
Experimental investigations of liquid metal convection become even more important once they are pushed to higher Rayleigh numbers in GaInSn or to even lower Prandtl numbers in liquid sodium.
In those parameter ranges, the total integration times in direct numerical simulations will be even shorter since the time-step width is ultimately limited by the strong temperature diffusion.
Given that in our case the thermal diffusion time has values of $170$ to $1300 \,\tau_\mathrm{ff}$, experiments of the present kind are currently the only way to study longer-term evolutions in these convection flows.

\bigskip
TZ and FS are supported by the Deutsche Forschungsgemeinschaft with Grants No. GRK 1567 and No. VO 2331/1-1, respectively.
We thank Janet D. Scheel and Christian Resagk for useful comments and discussions.

\appendix
\section{}\label{apx:PowerLawError}

In this work, power laws fits to data with errors on both, the abscissa and the ordinate, are conducted using orthogonal distance regression~\citep[ODR,][]{Boggs1990}.
To evaluate the accuracy of the resulting amplitude and exponent values, the fit is repeated multiple times while varying the measurement points randomly according to their uncertainties, which are assumed to be normally distributed.
This results in histograms of the amplitude and exponent, which converge to a constant distribution for a high enough number of repetitions.
The parameter errors are the standard deviations of these distributions with respect to the initial ODR fit results. 
The exponent histogram converges to a Gaussian distribution.
The amplitude, however, can be understood in terms of an inverse Gaussian distribution, which allows for positive values only. 
This procedure was adopted here in order to more accurately include the effect of the data uncertainty on the accuracy of scaling exponents.

\bibliographystyle{JFMtemplate/jfm}
\bibliography{literature}

\end{document}